\documentstyle[aps, epsf,graphicx]{revtex}
\newcommand{\AmS}{{\protect\the\textfont2
  A\kern-.1667em\lower.5ex\hbox{M}\kern-.125emS}}

\def\gsim{\stackrel{>}{\sim}}
\def\lsim{\stackrel{<}{\sim}}
\def\beq{\begin{equation}}
\def\eeq{\end{equation}}
\def\ba{\begin{array}}
\def\ea{\end{array}}
\def\ol{\overline}
\def\ul{\underline}

\def\half{\frac{1}{2}}

\def\eps{\epsilon}
\def\egzk{E_{\rm GZK}}
\def\dgzk{D_{\rm GZK}}
\def\tentwenty{10^{20}}

\def\dmpc{D_{\rm Mpc}}
\def\lammpc{\lambda_{\rm Mpc}}

\begin{document}

%\special{papersize=8.5in,11in}
\twocolumn
\renewcommand{\topfraction}{1.0}
\twocolumn[\hsize\textwidth\columnwidth\hsize\csname
@twocolumnfalse\endcsname
% declarations for front matter

%\title{North-south asymmetry at the high end of the cosmic ray spectrum?}
%\author{Luis A. Anchordoqui$^a$, Haim Goldberg$^a$, and Thomas J. Weiler$^b$}
%\address{$^a$Department of
%Physics, Northeastern University, Boston, MA 02115}
%\address{$^b$ Department of Physics \& Astronomy, Vanderbilt University,
%Nashville, TN 37235}

\title{Extreme-Energy Cosmic Rays: Puzzles, Models, and Maybe Neutrinos}
\author{Thomas J. Weiler}
\address{Department of Physics \& Astronomy,
 Vanderbilt University,  Nashville, TN 37235, USA}

\maketitle

\begin{abstract}

The observation of twenty cosmic-ray air-showers at and 
above $\tentwenty$~eV poses fascinating 
problems for particle astrophysics: 
how the primary particles are accelerated to these energies, 
how the primaries get here through the 2.7K microwave
background filling the Universe, and how the 
highest-energy events exhibit clustering on few-degree angular scales 
on the sky when 
charged particles are expected be bent by cosmic magnetic fields.
An overview of the puzzles is presented, followed by a brief 
discussion of many of the models proposed to solve these puzzles.
Emphasis is placed on (i) the signatures by which
cosmic ray experiments in the near future will discriminate 
among the many proposed models,
and (ii) the role neutrino primaries may play in 
resolving the observational issues.
It is an exciting prospect that highest-energy cosmic rays
may have already presented us with new physics 
not accessible in terrestrial accelerator searches.

\end{abstract}

\vskip2pc]

%
% typeset front matter (including abstract)
%
\section{INTRODUCTION}
An unsolved astrophysical mystery, now forty years old,
is the origin and nature of the extreme energy
cosmic ray primaries (EECRs) responsible for the observed events
at highest energies, $\sim\tentwenty$~eV \cite{EECRreviews}.
About twenty events at $\sim\tentwenty$~eV
have been observed by five different experiments \cite{Xpts}.
The origin of these events is a mystery,
for there are no visible source candiates
within 50 Mpc except possibly M87, a radio-loud AGN at $\sim 20$~Mpc,
and Cen-A (NGC5128), a radio galaxy at 3.4~Mpc,
and neither of these is in the direction of any of 
the observed events \cite{ES95}.
Since the observed events display a large-scale isotropy, many sources
rather than one source seem to be required.  The nature of the primary
particle is
also mysterious, because interactions with the 2.73K 
cosmic microwave background
(CMB) renders the Universe opaque to nucleons at 
$\egzk\sim 5\times 10^{19}$~eV,
and double pair production on the cosmic radio background (CRB) renders
the Universe opaque to photons at even lower energies.
The theoretical prediction of the end of transparency for nucleons at
$\egzk\sim 0.5\times \tentwenty$~eV is the famous 
``GZK cutoff'' \cite{gzk}.
Figure 1 shows a recent compilation of the AGASA 
data set, clearly extending beyond $\egzk$.

The main theory challenges in attempting to understand the 
super-GZK data are:
(i) What cosmic source could have accelerated the primary particles to
such extremely high energies?
and (ii) If the sources are distant ($\gsim 100$~Mpc), then 
how could their primaries have propagated through the
cosmic background radiation without substantial energy loss?
The acceleration mechanism either requires a Zevatron
accelerator ($1 ZeV\equiv 10^{21}$~eV) \cite{Zevatron},
distant because such a source could not be missed if it were nearby;
or speculative decaying super-massive particles (SMPs)
or topological defects (TDs) with mass-scale $\gsim 10^{22}$~eV,
clustered nearby;
or possibly magnetic monopoles accelerated by the cosmic magnetic fields.
From distant Zevatrons, only neutrinos among the known particles
can propagate unimpeded to earth.
Exploiting this fact are the ``Z-burst model,'' 
and the strongly-interacting neutrino model.
Adding to the drama and mystery 
at present is the observed {\bf large-scale isotropy}
and {\bf small-scale anisotropy}.  
Surely these characteristics hint at a solution to the mystery of origin;
protons, nuclei, and magnetic charges bend in cosmic magnetic fields,
whereas photons and neutrinos do not.
With more data above $\egzk$, several distinct telltale signatures 
including the isotropy/anisotropy will allow
one to discriminate among the many models proposed for the origin and nature
of the EECRs.  
\begin{figure}
\epsfxsize=80mm
\epsfbox{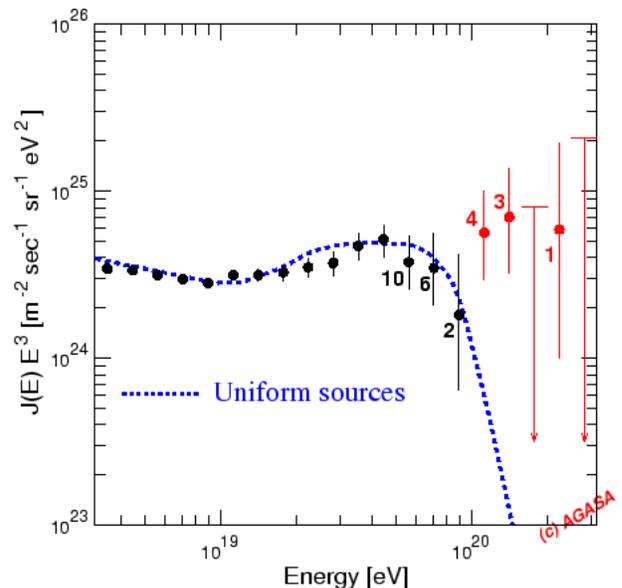}
\label{fig:AGASA}
\caption{ 
Extreme-energy cosmic ray 
spectrum as observed by AGASA. Error bars correspond
to 68 \% C.L. and the numbers count the events per energy bin.
The dashed line revealing the GZK cutoff 
is the spectrum expected from uniformly distributed 
astrophysical sources 
(from the AKENO website {\protect \cite{Xpt-websites}}).} 
\end{figure}

\section{EVENT CLUSTERING}
One revealing signature is already evident in the
existing data sample.  
This is the pairing of events on the celestial sky.
The AGASA experiment has already presented data strongly suggesting
   that directional pairing is occuring at higher
   than chance coincidence \cite{AGASApairs}.
Of the 47 published AGASA events above $4\times 10^{19}$~eV,
9 are contained in three doublets and one triplet with separation angle
less than the angular resolution of $2.5^{\circ}$.
The chance probability of this clustering occuring in an isotropic distribution
is less than 1\%.  The chance probability for the triplet alone is only 5\%.
Of the seven events above $10^{20}$~eV, three are counted among the doublet
events.
Most recently, AGASA has reported two more events 
above $\egzk$ \cite{teshima00}.
Each aligns in direction with a previous event,
reducing further the probability for random clustering to
0.07\% at 3$\sigma$.
Comparisons of event directions in a combined data sample of four experiments
further supports non-chance coincidences,
especially in the direction of the SuperGlactic Plane \cite{ALLpairs}.

Whether the pairing is random or dynamical,
we shall also know in the near future.
If the pairing turns out to be dynamical, I would argue that neutrinos are
a favorite candidate for the primary particles.
This is because photons have such a short ($\sim 10$~Mpc)
abosrption length, and protons are bent by cosmic magnetic fields during their
extragalactic journey.

If neutrinos are the primaries, they should point back to their sources,
thereby enabling point-source astronomy for the most energetic
sources of flux at and above $\tentwenty$~eV.
It was reported \cite{BF98} that the first five events at $\tentwenty$~eV
did in fact point toward extragalactic compact radio-loud quasars,
just the class of objects which could accelerate EECRs to 
ZeV energies via shock mechanisms.
With the inclusion of subsequent data, 
this association is controversial \cite{STAR00}.
The jury awaits further evidence.

\subsection{Coincidence or physics?}
\label{sec:coinorphy}
A random distribution is obtained by tossing
$n$ events randomly into
$N\simeq (\Omega/\pi\theta)^2
 = 1045\;(\Omega/1.0\ {\rm sr})(\theta/1.0^\circ)^{-2}$
angular bins,
where $\Omega$ is the solid angle
on the celestial sphere covered by the experiment
and $\theta$ is the bin half-angle.
Each resulting event distribution is specified by the partition of the $n$ total
events into a number $m_0$ of empty bins, a number $m_1$ of
single hits, etc., among the $N$ angular bins.
The probability to obtain a given event topology is \cite{GW00}
   \beq
   P(\{m_i\},n,N)=\frac{1}{N^n}\:
   \frac{N!\;\;n!}{\prod_{j=0} m_j!\;(j!)^{m_j}}\;.
   \label{eq:Hys}
   \eeq
The variables in the probability are not all independent,
as 
$\sum_{j=1} j\times m_j = n$ and
$\sum_{j=0} m_j = N$.
It is useful to use these constraints to rewrite this probability as
   \beq
   P(\{m_i\},n,N)=\frac{N!}{N^N}\,\frac{n!}{n^n}\,
   \prod_{j=0} \frac{(\ol{m_j})^{\,m_j}}{m_j!}\;,
  \label{eq:rewrite}
   \eeq
where we have defined
   \beq
   \ol{m_j}\equiv N\left(\frac{n}{N}\right)^j\frac{1}{j!}\;.
   \label{eq:meanm}
   \eeq
   %
%In the $n\ll N$~limit, $\ol{m_j}$
%is expected to approximate the mean number of $j$-plets,
%and eq. (\ref{eq:rewrite}) becomes roughly Poissonian.
%As an approximate mean, $\ol{m_j}$ defined in eq.\ (\ref{eq:meanm})
%provides a simple estimate of cluster probabilities due to chance
%for the $n\ll N$~case.

When $N\gg n \gg 1$, a limit valid for the
AGASA, HiRes, Auger and Telescope Array experiments,
one finds
   \beq
P(\{m_i\},n,N)\approx
\prod_{j=2} \frac{(\ol{m_j})^{m_j}}{m_j!} e^{-\ol{m_j}\,r^j (j-2)!}\,,
\label{eq:largeN}
   \eeq
where $r\equiv (N-m_0)/n\approx 1$.
%and a prefactor
%${\cal P}\equiv e^{-(n-m_1)}\,(n/m_1)^{m_1 +\half}\approx 1$
%has been omitted.
The non-Poisson nature of eq.\ (\ref{eq:largeN}) is reflected in the
factorials and powers of $r$ in the exponents.
%and the small deviation of the prefactor from unity.
%Note that $\ol{m_j}$ as defined in eq.\ (\ref{eq:meanm}) is
%the mean number of j-plets in a Poissonian sense.

%In Table 1 we list the typical parameters for each of the
%large-area experiments.
% \cite{Xpt-websites}.  
%
% Table of experiments
%   \begin{table}
%\begin{footnotesize}
%
%   \caption[] {
Typical values of effective area $A$~(km$^2$~sr),
celestial solid angle $\Omega$, and angular resolution
$\theta_{\rm min}$ for the existing and proposed EECR 
experiments \cite{Xpt-websites}
are shown in the following Table, where the 
incident flux $F(\ge \egzk)=10^{-19}{\rm cm^{-2} s^{-1} sr^{-1}}$
has been used to estimate the number of events ($n$/yr) 
above $\egzk=5\times 10^{19}$~eV.
   %
%   \vspace{0.5 cm}
\begin{center}
%\begin{footnotesize}
%\begin{small}
%   \centering\leavevmode
\begin{tabular}{|c|c|c|c|c|}
   \hline
   & AGASA & HiRes & Auger/TA & EUSO/OWL \\ \hline
   ${\rm km^2\,sr}$ & 150 & 800 & $6\times 10^3$ & $3\times 10^5$ \\
   \hline
   $n/{\rm yr}$ & 5 & 30 & 200 & $ 10^4$ \\
\hline
$\Omega\: ({\rm sr})$ & 4.8 & 7.3 & 4.8 & 4$\pi$ \\ \hline
   $\theta_{\rm min}$ & $3.0^\circ$ & $0.5^\circ$ & $ 1.0^\circ$ &
$1.0^\circ$ \\ \hline
%   $N=\frac{4}{\pi}\frac{\Omega}{\theta^2}$ &
%$4.2\times 10^4$ & $2.3\times 10^6$
%          & $3.8\times 10^5$& $6.6\times 10^5$
   \end{tabular}
%\end{footnotesize}
%\end{small}
\end{center}
%   \end{table}

\vspace{0.2 cm}

In Figure 2 are shown inclusive probabilities for the Auger experiment
with 100 events, as determined by formulae eq.\ (\ref{eq:largeN}).
Note that ``inclusive probability'' means the stated
number of $j$-plets {\it plus any other clusters}.
The 8-doublet probability is extremely sensitive to the angular
binning;
observation of a flatter dependence on angular bin-size could signal a
non-chance origin for the clustering.
The observation of two triplets with angular binning of less than $2^\circ$
would constitute 3-sigma evidence for cosmic dynamics.
The same conclusion would hold if a quadruplet within $2.5^\circ$
is observed among the first 100 Auger events.
\begin{center}
\includegraphics[height=2.5in]{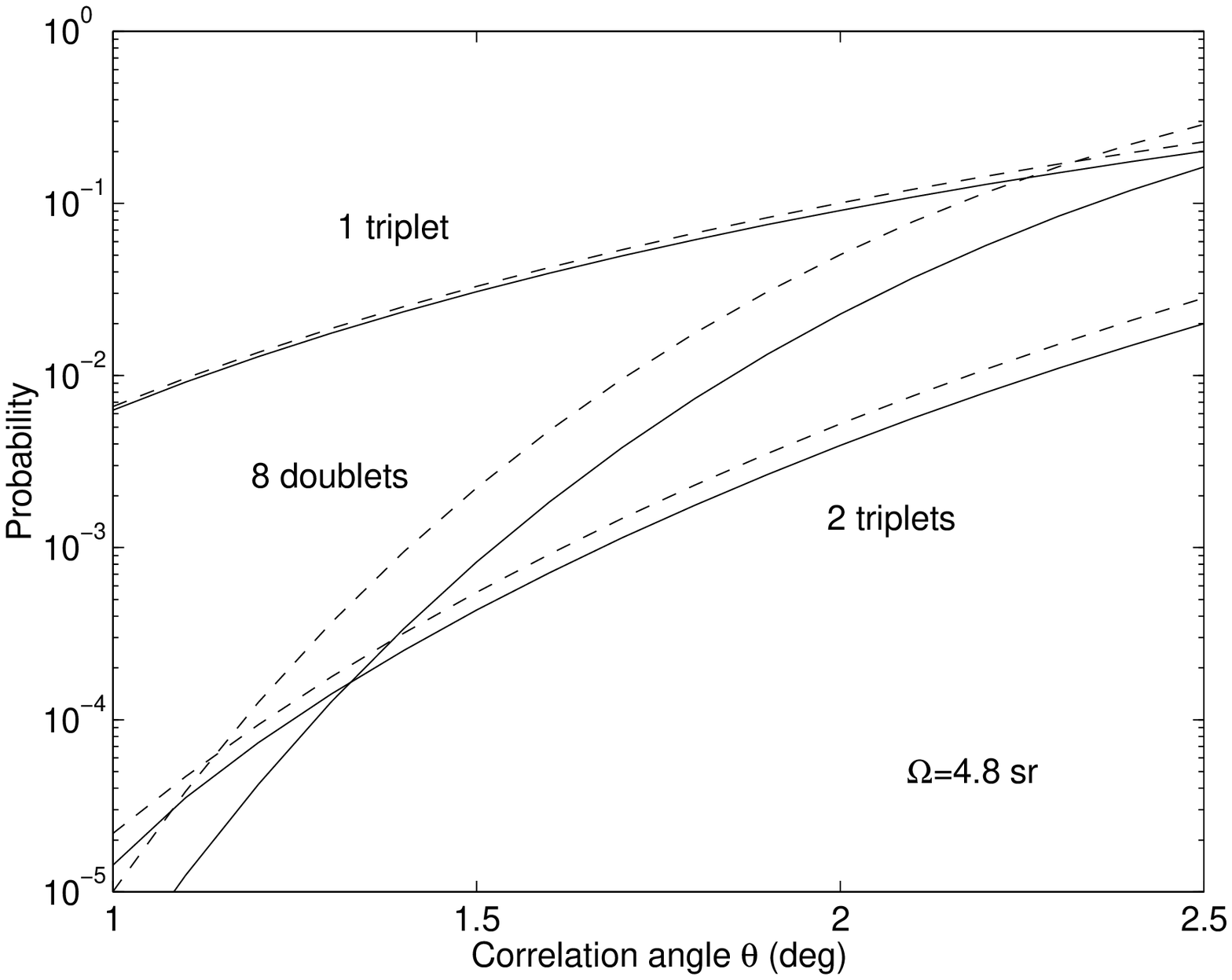}\\
\label{fig:Auger}
 \end{center}
%\caption{
{\small Figure 2: Inclusive probabilities for various clusters in a 100
event sample at Auger.
The solid line is the exact result, 
the dashed line is the Poisson approximation}\\
%\label{clustauger}
%
\vspace{0.3cm}

Without some modification, 
our analytic formula may not be directly relevant to HiRes data.
Clear, moonless nights are required for
detection of atmospheric fluorescence, 
and summer nights ($\sim 18$~hrs right ascension) are effectively 
40\% shorter than winter nights 
($\sim 6$~hrs right ascension) for monocular HiRes \cite{HRpc}.
Accordingly, the HiRes efficiency versus Galactic longitude 
varies significantly, roughly as 
$N(RA)={\overline N}(1+ \eps\sin(2\pi RA/24{\rm hrs}))$, 
with $\eps\approx 0.25$. 
One must ask whether this sinusoidally-varying efficiency 
invalidates the analytic approach with its assumption of a 
constant efficiency.
In the Monte Carlo approach there is an easy method to
generate ``background'' data sets with the experimental efficiency
properly incuded -- one randomly permutes the 
RA coordinates of the real data to remove any 
dynamical correlation among the events.
Fortunately, there is also a simple method to 
estimate the efficiency correction to an analytic Poisson distribution, if
one assumes that although $N$ and $n$ vary with right ascension,
$\xi\equiv n/N$ is constant.  Then it is easy to show that 
the Poisson distribution 
\beq
P(j,m) = \left(\frac{N\xi^j}{j!}\right)^m\frac{1}{m!}\,\,e^{-N\xi^j/j!}
\eeq
is corrected upwards by the factor 
\beq
{\cal I}=\frac{1}{2\pi}\int_0^{2\pi} d\phi (1+\eps \sin(\phi))^m 
\,\,e^{-{\overline N}\eps\sin(\phi)\xi^j/j!}\,.
\eeq
For $\frac{\eps \xi^{j-2}}{j!} \frac{n^2}{{\overline N}}\ll 1$,
the exponential is near unity and the correction has a closed form,
${\cal I}_m\approx (1-\eps^2)^{m/2} P_m (\frac{1}{\sqrt{1-\eps^2}})$, 
with $P_m$ a Legendre polynomial.  As a series, the correction is
\beq
{\cal I}_m = 2^{-m}\sum_{k=0}^{[m/2]} 
\frac{(-)^k (2m-2k)!}{k! (m-k)! (m-2k)!} (1-\eps^2)^k\,.
\eeq
In particular, for $n^2\ll {\overline N}$, we have 
${\cal I}_1 = 1$ independent of $\eps$, 
${\cal I}_2 = 1+\eps^2/2, = 1.03$ for HiRes,
${\cal I}_3 = 1+3\eps^2/2, = 1.09$ for HiRes, and 
${\cal I}_4 = 1+3\eps^2 +3\eps^4/8, = 1.189$ for HiRes.
The lesson is that HiRes efficiency corrections appear negligible
for small numbers of clusters,
and we may proceed with our analytic analysis for the HiRes experiment,
for which about 20 events at $10^{20}$~eV
are expected when the first full year's data is analyzed.
We display in Figure (3) the inclusive probabilities for
one or more, two or more, and three or more doublets;
   and one or more triplets, over a range of angular
   binning.
% using Eqs.~\req{Hys}and \req{bins}.
%
\begin{center}
\includegraphics[height=2.5in]{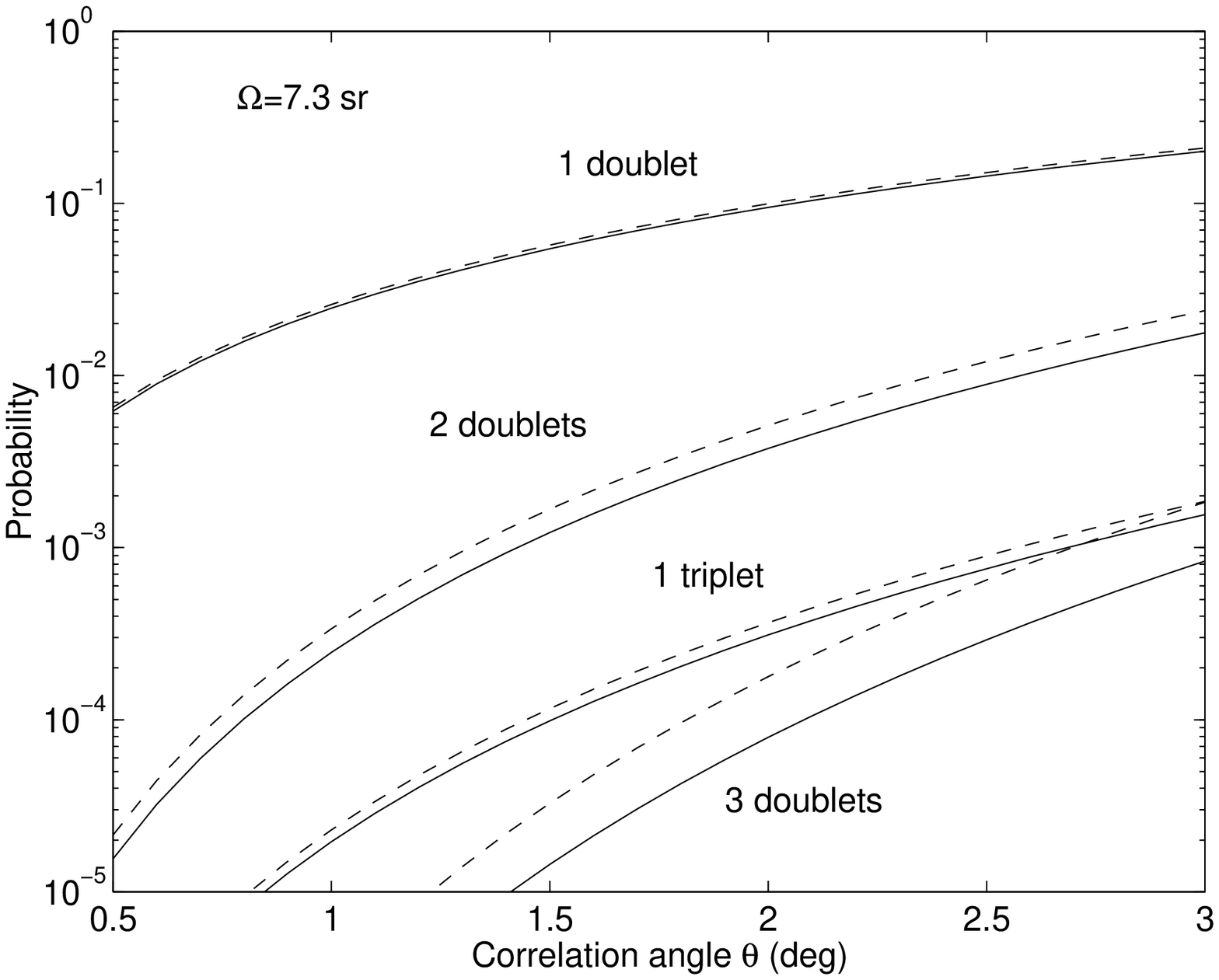}
\label{fig:HiRes}
\end{center}
{\small Figure 3: Inclusive probabilities for various clusters, given 20
events at HiRes.  The solid line is the exact result, 
the dashed line is the Poisson approximation.}
\vspace{0.3cm}

Note that for all except the 3 doublet configuration, the
   Poisson approximation using the mean values in Eq.(\ref{eq:meanm})
provides an estimate good to
   within 50\% of the non-approximate form;
for the (much suppressed) 3 doublet configuration,
it overestimates
the probability by about a factor of 3 in much of the angular region.
For angular binning tighter than $2^{\circ}$, an observation of two
   doublets among the first 20 events has a chance probability of less than
0.5\%.
   Thus the observation of this topology could be construed
   as evidence (at the 3$\sigma$ level) for clustering beyond statistical. The
   observation of a triplet within $\le 3^{\circ}$ has a random probability of less
   than $10^{-3}$, and hence observation of such a triplet would most
likely signify
   clustered or repeating sources, or magnetic focusing effects.
With the accumulation of 40 events (not shown in the figure),
the appearance of two
   doublets
   has a probability of less than 0.5\% for a correlation angle
   of $1^{\circ}$ or less. This illustrates how the good
   angular resolution of HiRes may be used to detect non-statistical
   clustering
   with only a few observed clusters.

Projected event rates for the EUSO/OWL/AW experiments present 
a pleasant problem for our analytic formula.
In the case where $n>N\gg 1$,
relevant for the EUSO/OWL/AW experiments after a year or more
of running, higher $j$-plets are common and
the distribution of clusters can be rather broad in $j$.
From eq.\ (\ref{eq:meanm}) we have
$(\ol{m_j}/\ol{m_{j-1}})=(n/jN)\sim (\pi n\theta^2/j\Omega)$.
Already at $j=1\,(2)$, Stirling's approximation to $j!$ is 
good to $8\%\,(4\%)$,
and so for $j\ge 1$ we may approximate
$\ol{m_j}\approx (N^3/2\pi en)^\half\,(en/jN)^{j+\half}$.
Extremizing this expression with respect to $j$, one learns that
The most populated $j$-plet occurs near $j\sim n/N$.
Combining this result with the broad distribution expected for large $n/N$,
one expects clusters with $j$ up to ${\rm several}\times \frac{n}{N}$
to be common in the EUSO/OWL/AW experiments.  Probabilities and
meaning for frequent, large clusters
are somewhat difficult to assess numerically.

The random distributions displayed in figs.\ (2) and 
(3) are
expected from some models, such as randomly-situated decaying 
super-massive particles (SMPs),
or charged-particle or magnetic monopole primaries with directions
randomized by incoherent cosmic magnetic fields.
A complementary approach to the random probabilities shown here
is to consider specific source models
generating non-random angular distributions. Steps along
this line of inquiry have recently been taken \cite{nonrandom}.
Future progress in the field will involve comparisons of the
random and non-random model predictions with the data.

\section{MODELS FOR SUPER-GZK EVENTS}
The conjectured origins of the super-GZK particles fall into
four basic categories.  These are
(i) nearby accelerators,
(ii) exotic primaries,
(iii) exotic physical law, and
(iv) neutrino primaries.
Other reviews \cite{EECRreviews} 
have emphasized the first three categories 
(especially (i) and (ii)),
so here we will be brief with those, and
put more emphasis on the neutrino-primary option.

\subsection{Nearby accelerators}
Several types of sources have been proposed to exist within our Galactic halo.
These include highly-ionized relativistic dust grains, Galactic supershocks,
young neutron stars, magnetars (highly magnetized pulsars),
decaying SMPs with GUT masses of order $10^{15}$~GeV 
or with inflation-motivated
masses of order $10^{13\pm1}$~GeV, 
topological defects \cite{TDold}
such as strings, Q-balls and vortons,
and annihilating monopole-antimonopole bound states (monopolonium)
\cite{mononium}.
For rare sources emitting charged particles, such as magnetars,
it is necessary to postulate that the primaries are iron nuclei to ease
the acceleration requirement and to isotropize the flux in our galactic
magnetic field. 
It is intriguing that the observed magnitude of 
CMB fluctuations fixes the reheat temperature following inflation
to $10^{13\pm1}$~GeV, which allows gravitational and thermal production
of TDS and/or SMPs of just this mass (if they exist) \cite{SMPreview};
this mass scale is just right for producing $\tentwenty$ 
secondary particles via decay.

For some TD models, dimensional arguments and scaling laws
seem to give an emission rate to short-lived SMPs consistent with 
the observed EECR rate without tuning exotic parameters.  
However, for most decaying sources such as SMPs, TDs, and monopolonium,
it is necessary to tune the lifetime to be longer but not too much longer 
than the age of the Universe in order to maintain an appreciable
secondary particle emission rate today.
Discrete gauged symmetries \cite{newsym} 
or hidden sectors \cite{crypton}
are introduced to stabilize the heavy particle.  
Then rather esoteric physics is needed to break the new
symmetry super-softly to maintain the long lifetime.  
High-dimension operators, 
wormholes \cite{wormhole}, 
and instantons \cite{instanton} 
bibitem{instanton} 
have been invoked for this purpose.

Models with sources mainly in the Galactic plane are disfavored
by the lack of any observed planar asisotropy in the data.
Models with sources clustered in the Galactic halo predict a dipole
enhancement in the direction of the Galactic center \cite{dipole},
which will be tested by the Auger Observatory in a few years.

Possible sources outside our halo, but still relatively nearby, include 
the radio-loud quasar M87 at 18 Mpc \cite{M87}, 
the similar Cen-A at 3.4 Mpc \cite{CenA},
rare nearby GRBs, and now-dormant rare AGNs
(also called massive dark objects -- MDOs).  
These sources are few at best, and strong
magnetic fields must be postulated to isotropize and/or 
confine their emissions.
In the case of GRBs, the identification of the red-shift 
of their host galaxies at typically $\gsim 1$ renders a local
occurence of a GRB highly improbable \cite{noGRB}.

Finally, primordial black holes (PBHs) have also been suggested,
but any sensible initial mass spectrum is unable to provide 
a sufficient number of PBHs in the final stage of decay today.

\subsection{Exotic primaries}
The GZK cutoff can be raised by simply postulating a primary hadron
slightly heavier than the proton.  The reason is kinematical -- the
cutoff energy varies as the square of the mass of the first excited
resonant state. For this reason, Farrar has proposed that
light supersymmetric baryons, made from a light gluino plus the
usual quarks and gluons, may be the primary EECRs \cite{SUSYlite}.  
Such a scenario
renders the source-energetics issue even more challenging.
In any event, terrestrial experiments seem to have recently 
closed the window on a possible light gluino. 
%\cite{noSUSYlite}.

Another interesting possibility for the primary EECR
is the {\bf magnetic monopole} \cite{PorterGoto,KW96}.
Any breaking of a semisimple gauge symmetry
which occurs after inflation and which
leaves unbroken a $U(1)$ symmetry group may produce an
observable abundance of magnetic monopoles.
The monopole mass is expected to be
$\sim \alpha^{-1}$ times the temperature $T_c$ 
of the symmetry breaking.
At the time of the phase transition, roughly one monopole
or antimonopole is produced per correlated volume.
The resulting monopole number density today is
\beq
n_M \sim 10^{-19}\, (T_c/10^{11}{\rm GeV})^3 (l_H/\xi_c)^3\,{\rm cm}^{-3},
\label{monodensity}
\eeq
where $\xi_c$ is the phase transition correlation length,
bounded from above by the horizon size $l_H$ at the time of
the transition.
In a second order or weakly first order phase transition,
the correlation length is comparable to the horizon size.
In a strongly first order transition,
the correlation length is considerably smaller than the horizon size.

The kinetic energy imparted to a magnetic monopole on traversing a magnetic
field $B$ is $E_K\sim g B \xi$,
where $g=e/2\alpha$
is the magnetic charge according to the Dirac quantization condition,
and $\xi$ specifies the
field's coherence length.
Given the magnitude and coherence length data for the cosmic magnetic fields,
monopole kinetic energies in the range $10^{20}$ to $10^{23}$~eV are
expected; the acceleration problem is naturally solved.
Monopoles with $M\lsim 10^{14}$~GeV
should be relativistic, and carry the appropriate energy
to qualify as candidates for the EECR primaries \cite{KW96,WKWB00}.
Within field theory there exist many possibilities for
an intermediate unification scale and intermediate-mass monopoles.

The propagation problem is also naturally solved.
The scattering cross-section for the monopole on the 3K and diffuse
photon backgrounds is just classical Thomson, 
valid even for strong coupling:
$\sigma_{\rm T}=8\pi\alpha_M^2/3M^2
\sim 2\times 10^{-43}\,(M/10^{10}{\rm GeV})^{-2}\,{\rm cm}^2$.
The resulting mean free path for inverse Compton scattering is many
orders of magnitude larger than the Hubble size of the Universe.

The relativistic monopole flux is simply
\beq
F_M=c\,n_M/4\pi \sim 2\times 10^{-19}\,(M/10^{10}{\rm GeV})^3 (l_H/\xi_c)^3
\label{monoflux}
\eeq
per cm$^2$-s-sr,
which compares favorably with the integrated flux above $10^{20}$~eV,
$F_{\rm data}(>10^{20}{\rm eV})\sim 2\times 10^{-20}
{\rm cm}^{-2}{\rm s}^{-1}{\rm sr}^{-1}$,
and is comfortably below Parker's upper bound 
$F_{\rm Parker}=10^{-15}{\rm cm}^{-2}{\rm s}^{-1}{\rm sr}^{-1}$ 
for a cosmic monopole flux.

Signatures for EECR monopoles in our atmosphere and in ice
are discussed at length in \cite{WKWB00}.
Of particular interest as a model for the super-GZK primaries
is the ``baryonic monopole'' \cite{barymono,WKWB00}.
It is a bound state of three colored monopoles,
confined by chromomagntic strings.  Chromomagnetic strings can
stretch when excited, but cannot break into quark-antiquark pairs.
On first interaction in the atmosphere, the baryonic monopole 
of mass $M$ and energy $E=\gamma M$ stretches to create a 
huge geometrical cross-section of order 
$\gamma/\Lambda^2_{\rm QCD}\sim (\gamma/10^6)\times 10^7\,$~mb.
Consequently, nearly all of the initial monopole energy is transferred
to the atmospheric shower in a very short distance.
A recent simulation of the baryonic-monopole
showed good agreement with the lateral muon and hadron content of the
highest-energy Yakutsk event, but less than good
agreement with the longitudinal profile 
of the Fly's Eye event \cite{OKdokey00}.

Any confirmed directional pairing of events would appear difficult
to achieve with the monopole model.
Also, in the context of a model of the Galactic magnetic field, 
it has been shown that some memory of the local 
spiral arm direction, and an energy spectrum 
flatter than the observed one, are expected in the data 
if monopoles are the primaries \cite{noMono}.
Although the directional criterion  may not survive inclusion of
extragalactic fields, the flatness of the spectrum probably does.

\subsection{Exotic physical laws}
The most remarkable proposals posit a breakdown of Lorentz
invariance \cite{LIV}
or a breakdown of general relativity above some high scale.
In the case of broken GR, it could be 
spacetime fluctuations (expected in a theory of quantum gravity)
wiggling the on-shell dispersion relation \cite{qGrav} 
or appearing as $1/M_P$ operators
which alter the physics.
String theory provides motivation for GR-breaking at a
possibly lower scale, the string scale $M_S$.
%\cite{qGrav}.
Certainly the energy window of EECRs 
($\tentwenty {\rm eV}/M_P\sim 10^{-8}$)
is beyond that of terrestrial
accelerators, and so ripe for 
speculation on new high-energy physics.

Signatures of models with no photo-pion production above $\egzk$
include the absence of a proton pile-up below $\egzk$,
the absence of a cosmogenic neutrino flux, and possibly
undeflected pointing 
of the primary back to its source.

\subsection{Neutrino primaries}
\label{sec:nuprimary}
Turning to the possibility that the primaries may be neutrinos,
one encounters an immediate obstacle:
the SM neutrino cross-section
is down from that of an electromagnetic or hadronic interaction by six
orders of magnitude. This implies a low air-shower 
rate, and an accumulation of events
at low altitudes (``penetrating'' events) where the target density is highest.
On the other hand, the neutrino-primary hypothesis is supported 
by the observed clustering discussed earlier. 
Two solutions to the small cross-section problem for
primary neutrinos have been proposed.

\subsubsection{Neutrino annihilation to $Z$--bursts}
Here it is proposed that the primary particles which propagate across cosmic
distances above the GZK cutoff energy are neutrinos, which then annihilate
with the cosmic neutrino background (CNB) 
within the GZK zone ($D< D_{\rm GZK}\sim 50$~Mpc)
to create a ``local''
flux of nucleons and photons above $\egzk$,
as shown in Fig. (4).
It was noted many years ago that 
a cosmic ray neutrino arriving at earth from a
cosmically distant source has an annihilation probability 
on the relic--neutrino background of roughly $3.0\,h^{-1}_{65}\,\%$
(neglecting cosmic expansion) \cite{W82}. 
The probability for a neutrino with resonant energy to annihilate to a
$Z$-burst within distance $\dgzk$ is then  
$2.5\times 10^{-4}$ for a homogeneous CNB \cite{W97}.
The annihilation rate depends
upon the CNB density, reliably predicted 
in the mean  by Big Bang cosmology,
and on the Standard Model (SM) of particle physics.
The local annihilation rate 
is larger if our matter-rich portion of the Universe clusters
neutrinos \cite{W97,FMS97}, or if there is an intrinsic CP-violating 
$\nu-{\bar \nu}$ asymmetry \cite{GK99}.

\begin{center}
%\includegraphics[height=4.2in,width=4.8in]{nu2.eps}
%\vspace*{1cm}
\includegraphics[height=3in]{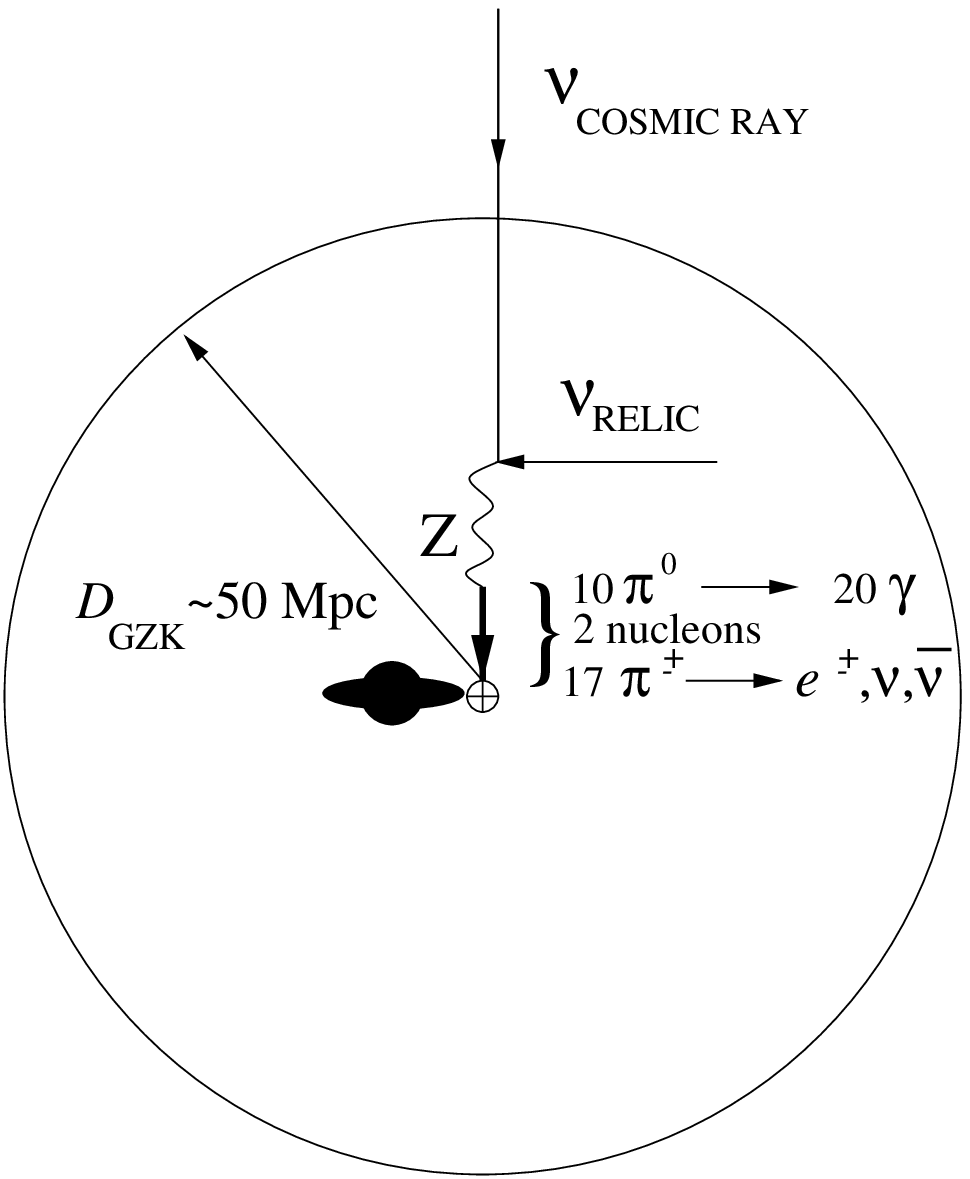}\\
\end{center}
%\caption{
{\small Figure 4: Schematic diagram showing the production of a $Z$--burst
resulting from the resonant annihilation of a cosmic--ray neutrino on a
relic (anti)neutrino.  If the $Z$--burst occurs within the GZK zone
($\sim 50$~to~100~Mpc) and is directed towards the earth,
then photons and nucleons with energy above the GZK cutoff
may arrive at earth and initiate super--GZK air--showers.}
\label{fig:schematic}
\\

Each resonant neutrino annihilation produces a $Z$ boson
%which immediately decays (its lifetime is
%$3\times 10^{-25}\;{\rm s}$ in its rest frame).
with a 70\% branching ratio into hadrons
known to include on average about one baryon--antibaryon pair,
seventeen charged pions, and ten neutral pions \cite{RPP}.
The ten $\pi^0$'s decay to produce
twenty high--energy photons.
For $m_\nu$ in the range $\sim$~0.1 to 2~eV,
the energy in this resonant ``$Z$-burst'' is 
fortuitously situated sufficiently above $\egzk$ at
\beq
E_{\nu}^R = M_Z^2/2m_{\nu} =
4\,({\rm eV}/m_{\nu}) \times 10^{21} {\rm eV}
\label{eqn:Eres}
\eeq
so as to produce photons and nucleons with energies exceeding
$\egzk$.\footnote
{
The resonant-energy width is narrow, 
reflecting the narrow width of the Z-boson: at FWHM 
$\Delta E_R/E_R \sim\Gamma_Z/M_Z = 3\%$.
}
The mean energies of the $\sim 2$~baryons and $\sim 20$~photons
produced in the Z decay are easily estimated.
Distributing the Z-burst energy among the mean multiplicity 
of 30 secondaries in Z-decay \cite{RPP},
one has 
\beq
 \langle E_p \rangle \sim \frac{E_R}{30} 
\sim 1.3 \left(\frac{{\rm eV}}{m_j}\right)\times 10^{20}{\rm eV}\,.
\eeq
The photon energy is further reduced by an additional factor of 2 
to account for their origin in two-body $\pi^0$ decay:
\beq
 \langle E_{\gamma} \rangle \sim \frac{E_R}{60} 
\sim 0.7 \left(\frac{{\rm eV}}{m_j}\right)\times 10^{20}{\rm eV}\,.
\eeq
Even allowing for energy fluctuations about mean values, 
it is clear that in the Z-burst model the relevant
neutrino mass cannot exceed $\sim 2$~eV.
On the other hand, the neutrino mass cannot be too light
or the predicted primary energies will exceed the observed
event energies.
In this way,
one obtains the approximate 0.1~eV lower limit on the
neutrino mass,
when allowance is made for an order of magnitude energy-loss 
for those secondaries traversing 50 to 100 Mpc. 

The challenging issue of how experiments might 
actually determine the absolute 
neutrino mass is discussed in \cite{PW01}.

If the $Z$--burst points in the direction of earth and occurs
within the GZK distance, then one or more of the photons and nucleons in the
burst may initiate a super--GZK air--shower at earth \cite{W97,FMS97}.
For a sufficient cosmic neutrino flux,
the hypothesis successfully explains the observed air--showers above $\egzk$.
Comparisons of the model predictions to super-GZK data
are available in \cite{Zsimns}.

The existence of neutrino mass in the desired range
seems nearly guaranteed from the tritium decay upper bound
\cite{BWW98}
and the lower bounds inferred from the terrestrial
neutrino oscillation experiments.
The simplest explanation for the atmospheric neutrino results
is neutrino oscillations driven by a mass--squared difference of
$\delta m_{\rm atm}^2 \sim 3\times 10^{-3}{\rm eV}^2$ \cite{atmosc},
which implies a neutrino mass of {\it at least} 0.05 eV.
Also, the recent LSND measurement appears to indicate a
mass--squared difference $\delta m_{\rm LSND}^2 \gsim
0.2 {\rm eV}^2$ \cite{LSND},
from which one deduces a neutrino mass of at least 0.5 eV.
%
%We summarize the neutrino mass lower bounds,
%and the corresponding $Z$-burst energy upper bounds,
%in Table 1.\\ HELP: TABLE HERE
%
From these lower bounds on neutrino mass,
one gets upper bounds on the $Z$--burst energy of
$10^{23}$ and $10^{22}$ eV, respectively, 
just right for extending the air-shower spectrum an order of magnitude
or two beyond the GZK cutoff!

A considerable cosmic neutrino flux above $\egzk$ is required for the
$Z$--burst hypothesis to successfully explain the super--GZK events.
The requirement is that
the product of the resonant energy times the 
neutrino flux at the resonant energy per flavor,
times the annihilation probability within the GZK zone
(which may be as large as 0.025\% to 1\% due to neutrino clustering),
times the photon and nucleon multiplicity per burst ($\sim$~20), 
is comparable to the observed flux at $\tentwenty$~eV.
The resulting requirement on the neutrino flux is roughly
$E_R\,F_{\nu_j}(E_R)\sim 10^{-18.5\pm 1}$/cm$^2$/s/sr.
Such a  neutrino flux at
$E \sim 10^{22}$ eV is directly measureable 
in a teraton ($10^{12}$~ton) detector
like EUSO/OWL/AirWatch,
and possibly in a search
for radio pulses produced by high energy neutrinos
penetrating a small column--density of matter in
the limb of the moon \cite{GLN99}.

While certainly large, this required 
neutrino flux violates no existing limits.
It has been pointed out \cite{LargeFlux}
that this flux cannot extrapolate as $E^{-2}$ to 
$10^{17}$~eV, for then it would violate the Fly's Eye bound 
arising from nonobservation of penetrating horizonal 
(i.e., neutrino induced) air showers at that energy.
It has also been pointed out \cite{EGRETprob}
that local neutrino clustering
is required to avoid generating a density of 30~MeV to 
100~GeV photons in excess of the EGRET 
experimental bound \cite{EGRET},
from distant Z-bursts undergoing electromagnetic cascading.
The extreme-energy neutrino flux implied by the $Z$-burst model 
probably requires unusual source dynamics \cite{CDF00}.
Among the reasons to hope that Nature obliges is that 
resonant neutrino annihilation provides the best hope at present
to actually measure the relic neutrino density.

\subsubsection{Strong $\nu$ Cross-section at $\gsim\egzk$}
It is interesting and suggestive that
the observed EECR flux beyond $\egzk$ is well matched
by the flux predicted for cosmogenic neutrinos.
This is not a complete coincidence. With the GZK cutoff,
any continued nucleon flux beyond $\egzk$ is degraded in energy,
photo-producing pions which in turn decay to produce cosmogenic neutrinos.
The number of produced neutrinos compensates for their lesser energy,
with the result that the neutrino flux matches well
to the observed super-GZK flux. One may entertain the notion that
the cosmogenic neutrinos \ul{are} the super-GZK primaries,
and that these neutrinos acquire a strong cross-section at $\sim 10^{20}$~eV.

Limits on the strength of
the neutrino cross-section at $\tentwenty$~eV can be inferred from
existing data.  Heuristically, one argues as follows.
The GZK process ensures that there is a flux of cosmogenic
neutrinos at $\tentwenty$~eV, with an easily calculated flux.
If the neutrino cross-section were weak,
an experiment looking for penetrating air-showers initiated
by the cosmogenic neutrinos would see nothing.
If the cross-section were strong
enough, the neutrino could not penetrate the atmosphere at all.
So the fact that the Fly's Eye experiment saw no penetrating showers
tells us that the neutrino cross-section is either strong or weak;
the mid-range is excluded.  The vertical depth of our atmosphere is
$x_{\rm v}= 1033\,{\rm g/cm}^2$, and the horizontal depth $x_{\rm h}$
is about 36 times greater.  In terms of the mean free path 
$\lambda$ of a
particle with cross-section $\sigma$, one has
$x_{\rm v}/\lambda = \sigma/1.6 {\rm mb}$, 
and $x_{\rm h}/\lambda = \sigma/44\mu{\rm b}$.
Thus, an estimate of the excluded cross-section is
$\sim 40\mu{\rm b}$ to $1 {\rm mb}$.
A more careful calculation has been performed, with the
result that $\sim 20\mu{\rm b}$ to $1 {\rm mb}$ is excluded \cite{TOS00}.
Hypothetical high-energy neutrino cross-sections 
in excess of a mb remain viable.

The idea that neutrinos, indeed, all particles, may have a strong
interaction at a high but observable energy scale is not new \cite{oldSInu}.
However, some recent ideas concerning new interactions 
relate well to the $\tentwenty$~eV scale.
One idea is that leptons are bound states of dual QCD gluons,
which reveal themselves just above the electroweak (EW) scale
at parton-parton $\sqrt{s}\sim$~TeV \cite{dualQCD}.
Another idea is that
grand unification occurs precociously at $\sqrt{s}\sim$~TeV,
because of extra dimensions or other reasons, and a neutrino
above this threshold becomes strongly-interacting via leptoquark
resonances \cite{Domokos}.
A third idea is that the exchange of a towers of Kaluza-Klein (KK)
modes from extra compactified dimensions lead to a strong
neutrino cross-section above $\sqrt{s}\sim$~TeV \cite{nuKK}.
In all three cases, it is the combination of a low $\sim$~TeV scale
for radically new physics and a quickly rising spectrum
of new states
(possibly increasing exponentially, $\rho\sim e^{\sqrt{s/s_0}}$)
that provides a rapid turn-on of a strong cross-section
for the neutrino.
Through unitarity, the new threshold at $\sim$~TeV has
consequences for cross-sections 
at lower energies \cite{GW99}, but they are not dramatic.
%as would be expected in single-particle exchange models \cite{BGH97}.

The KK exchange model may fail \cite{KP99} in that the
KK modes couple to neutral currents, and the scattered neutrino
carries away 90\% of the incident energy per interaction,
thereby elongating the shower profile.
But if the neutrino cross-section can be made large enough,
$\gsim 20$~mb, 
then multiple scattering within a nucleus 
will effect a sufficiently large energy 
transfer and save the model \cite{Jain2}.
Independent of the neutral current issue, 
the dual QCD and TeV-scale unification models seem to
provide viable explanations of the super-GZK data.
However, a recent calculation for the rate of rise of the 
low-scale unification cross-section in a string context
is not encouraging \cite{noexpo}.

Signatures for these models include directional 
pointing back to the EECR source, 
longitudinal shower profiles differing somewhat from
those of a proton or a gamma, and 
a strong correlation between
observed energies and zenith angle.
The latter signature should show an inverse proportionality
between the neutrino-air interaction length and the rising (with energy) 
neutrino cross-section.

\section{MODEL SIGNATURES}
There are several telltale discriminators to be sought
in higher statistics data.
These will eventually eliminate most (perhaps all!) of the models
so far proposed for the super-GZK events.
We list some discriminating signatures and discuss them.

\subsection{Small-scale anisotropies and pointing}
The discriminatory power of small-scale clustering was discussed
already in \S \ref{sec:coinorphy}.
Here we add some detail to the discussion.
In traversing a distance
$D$, a charged particle interacting with magnetic domains having coherence
length $\lambda$ will bend through an energy-dependent angle\footnote {On
average, half of the interactions of a super-GZK nucleon with the CMB change
the isospin.  At energies for which $c\tau$ of the neutron is small compared to
the interaction mfp of $\sim 6$~ Mpc, the neutron decays back to a proton with
negligible energy loss and the bending-angle formula is unchanged. However, at
the energy $6\times
\tentwenty$, $c\tau$ for the neutron is comparable to the interaction mfp, so
at higher energies the nucleon bending-angle is reduced by $\lsim 2$. }
\beq
\delta\theta\sim 0.5^\circ
\times \frac{Z\,B_{nG}}{E_{20}}\sqrt{\dmpc\,\lammpc}\,.
\eeq
\label{eq:bending}
Here $B_{nG}$ is the magnetic field in units of nanogauss, $E_{20}$ and $Z$ are
the particle energy in units of $10^{20}$~eV and charge, and the lengths $D$
and $\lambda$ are given in units of Mpc. It is thought likely that coherent
extragalactic fields are nanogauss in magnitude \cite{nG},
in which case super-GZK primaries from $\lsim 50$~Mpc will typically 
bend only a few degrees 
(but note that protons at $10^{19}$~eV will bend through $\sim
30^\circ$).
Thus, local models either postulate many invisible sources
isotropically-distributed with respect to the Galaxy to provide the roughly
isotropic flux observed above $\egzk$, or postulate a large extragalactic
magnetic field 
to isotropize over our Northern Hemisphere the highest-energy
particles from a small number of sources \cite{CKB01}.
Among the latter category,
some models postulate helium or iron nuclei as the
primaries, to increase the bending by the charge factors 2 and 26,
respectively.
For those models invoking
randomly distributed, decaying super-massive particles (SMPs) \cite{SMPs}
or topological defects (TDs) \cite{TDs} as sources,
and models invoking a large magnetic field with considerable
incoherent component, one may expect a nearly chance distribution of
observed events on the sky.  
However, there may be some clustering even in these models, due to
possible small-scale density fluctuations in the local SMP or TD distributions
\cite{locfluc}, or due to possible caustics in the
projection of large-scale extragalactic magnetic
fields on our sky (assuming the incoherent
magnetic fields are sufficiently small) \cite{caustics}.
In the SMP and TD models, a high photon fraction in the primary composition
further enhances clustering possibilities.

From the point of view of opening a new window to astronomy,
those models in which the primaries do point back to their
distant, active sources are the most interesting.
These models are few in number.
They are the Z-burst model, the strongly-interacting neutrino model,
and the quantum-gravity/LI-violating models.

\subsection{Large-scale anisotropies}
On large scales, one seeks associations of the CR directions with
the Galactic halo (to be revealed by a dipole anisotropy
favoring the direction of the Galactic center) or the local galactic
magnetic field,
with matter distributions in nearby galactic or super-galactic
clusters such Virgo, or with possible large coherent galactic or
extragalactic magnetic fields \cite{largeanisots}.
For large-scale studies, the Southern hemisphere Auger experiment will
prove invaluable for several reasons.
It offers coverage of potential sources and matter distributions,
and galactic and extragalactic magnetic fields,
not available from the North.
Moreover, it offers a view of our Galactic center which will provide
a North-South dipole discriminator
for or against a halo-centered population of sources such as
magnetars or halo-bound SMPs.
Southern Auger will also discriminate the M87-source model \cite{M87}
wherein EECRs are channeled by a hypothesized galactic magnetic wind
into the Northern hemisphere, and the Cen-A source model \cite{CenA}
which also yields a dipole anisotropy.
Of course, an orbiting experiment with $4\pi$ vision
like EUSO/OWL will be an even better instrument
for multipole analysis.

\subsection{Energy-direction-time correlations}
Because bending of charged particle trajectories
by intervening magnetic fields increases as particle energy decreases,
one may learn about the strength and geometry of extragalactic fields
from relative time delays and angular correlations 
of particles from a common source.
%I have misplaced my \cite{Etangle} reference, aminly to Sigl et al.
One may also learn about the source.
Quantitatively, the increase in path length due to bending 
leads to a relative
   increase in travel time of $\delta t/t\sim (\delta\theta)^2$,
   for small bending angle.
   Adding the contributions from the coherent magnetic domains then yields
\beq
\delta t\sim 300\,D_{\rm Mpc}\,
\left(\frac{Z\,B_{nG}\lambda}{E_{20}}\right)^2\,{\rm yrs}\,
\label{eq:tEcorln}
\eeq
for the time delay.
The time separation at earth is obtained by taking differences in eq.
(\ref{eq:tEcorln});
to first order in $\delta E$ it is already large:
\beq
t_1-t_2 \sim 600\,\dmpc\,
 \left(\frac{Z\,B_{nG}\lambda}{E_{20}}\right)^2\,
\left(\frac{\delta E}{E_{20}}\right)\,{\rm yrs}\,.
\label{eq:deltat}
\eeq
The correlation in energy and time becomes even more significant
   when it is remembered that the higher energy primary has an even higher
   mean energy in transit, before losses on the 2.7K background.

Surprisingly,
one AGASA event-pair has the higher-energy $1.06\times\tentwenty$ 
primary arriving
about 3 years {\sl after} a $0.44\times\tentwenty$ primary.
Assuming these primaries originate form a common source,
a possible explanation is that their source has a duration of at least
$3/(1+z)$~years (the red-shift factor is due to time-dilation).
Such a long-lived source does not occur in one-time burst
models (e.g. GRBs).  It may occur in 
decay models (e.g. SMPs, TDs, monopolonium)
if sub-clustering exists on small-angular scales withing halo clusters 
\cite{DKanisot}.
A recent paper \cite{KK00} 
notes that such sub-clusters may actually be observable
as micro-lenses for stars and background galaxies.
Counter-intuitive reverse pairing with the 
{\sl earlier} arrival time for the lower
energy charged-primary can also occur in 
certain magnetic field configurations, as shown in \cite{earlier}.

\subsection{Composition of the primaries}
Another signature to be sought is a statistical identification of
the nature of the primaries as a function of their energies.
Three methods have been identified to distinguish photon-initiated
showers from hadron-initiated showers.
One method relies on the longitudinal profile of the event,
particularly the depth at shower maximum $x_{\rm max}$.
The longitudinal profile of the Fly's Eye event at $3\times\tentwenty$~eV
(well-measured by its nitrogen fluorescence trail)
is ill-fit by a photon primary, well-fit by an iron primary, 
and somewhat fit by a proton primary \cite{gaisserMex}.
The second method relies on measurement of muon number, with a high
muon number purportedly favoring a hadron primary over a photon, 
and vice versa.
A recent study of the muon content of showers above $10^{19}$~eV 
seems to favor nucleons over photons \cite{Ave}.
However, caution is warranted with this method, in that 
some simulations show little difference in the muon-content
of showers from nucleon primaries 
versus photon primaries \cite{Aharonian};
and photons have themselves a 
significant partonic component at high energy.
%\cite{phoparton}.  \bibitem{phoparton}
Nevertheless, it is safe to say that photon primaries appear 
disfavored at present, but more data is needed before conclusions are drawn.

The third method of gamma identification will 
will rely on a predicted characteristic N-S vs.\ E-W gamma asymmetry.
This quadrupole asymmetry results from the polarization-dependent 
interaction of the gammas with the earth's magnetic field.

It is interesting to mention that the highest-energy 
Yakutsk event has an unusually high muon number;
the only model so far which successfully explains this 
invokes a magnetic baryonic-monopole as the primary particle 
\cite{OKdokey00}.
Basically, a relativistic baryonic-monopole of mass $M$ 
and incident energy $E$
showers like a giant nucleus of atomic number 
$A_{\rm eff}\sim M/\Lambda_{\rm QCD}\sim 10^6\,(M/{\rm PeV})$
and the same $\gamma=E/M$.
Intermediate mass monopoles therefore generate 
many, many charged pions which decay to muons.

All models wherein the primaries arise from QCD jets produce many more pions
than baryons.  The neutral pions in turn produce gammas. In these models,
the ratio of gammas to baryons is typically of order ten at the source.
Even allowing for the shorter attenuation length of photons 
relative to nucleons,
the measurement of the primary compositon for super-GZK events
becomes an excellent discriminator for models with jet-producing sources.
These models include $Z$-bursts, and decaying SMPs and TDs.

\subsection{Possible $E_{\rm max}$ energy cutoff}
The predicted of a cutoff at $\egzk$ is wrong.
Does Nature provide an alternative cutoff
within our reach?
Or do the data continue beyond our reach?
The shock-jock experts claim that it is difficult for conventional
shock-acceleration mechanisms to produce ZeV proton energies;
for this class of model, an $E_{\rm max}$ below $10^{22}$~eV
is certainly expected \cite{shockjock}.
Decaying SMP models also have a natural cutoff, at half of the SMP mass.
This could be as high as $E_{\rm max}\sim 10^{24}$~eV
for a long-lived GUT-mass particle,
but would be lower for other postulated SMPs.
In the $Z$-burst model there is a natural cutoff related to the tiny mass of
neutrinos: $E_{\rm max} = 4\,({\rm eV}/m_\nu)$~ZeV\@.
Implications from the atmospheric SK data
are that this cutoff is at most $7\times 10^{22}$~eV.

\subsection{CR flux above vs. below $\egzk$}
A ``smoothness'' variable such as
$R_j\equiv F_j (E>\egzk)/F_j (E<\egzk)$ for each
primary species $j$=nucleon, photon, iron nuclei, neutrino, etc.\ may be
revealing \cite{smoothness}.  
For primaries with a GZK cutoff, 
$F(E<\egzk)$ samples sources from the
whole volume of the Universe, and may even include
cascade products from $F(E>\egzk)$, 
whereas $F(E>\egzk)$ samples 
just the GZK volume; 
for primaries without a GZK cutoff,
$F(E<\egzk)$ and $F(E>\egzk)$ sample sources from the
whole volume of the Universe.
Lumps, bumps, and gaps in the spectrum near
$\egzk$ are a consequence of some models.
For example, hadron and neutrino pile-ups just below $\egzk$
are expected from the photo-pion production process which
occurs above $\egzk$ \cite{pileup}.  
Other sources of neutrino pile-ups
have also been suggested \cite{W97,KP00}.
Present data show continuity.
Smoothness studies of various models require simulation,
and are just beginning.

\subsection{Spectral index above $\egzk$}
One means of achieving more events above $\egzk$ is to postulate
a flattening of the primary proton spectrum at highest energies.
With more data, the extreme-energy spectrum will be measured.
A flattening of the $E^{-2.7}$ power law inferred from just
below $\egzk$ would indicate and constrain new sources.

\subsection{Measureable neutrino flux above $\egzk$}
If there is a new source of primary neutrinos above $\egzk$,
or if the neutrino cross-section becomes strong at super-GZK energies,
then there is the possibility that the primary neutrino flux
can be measured in an EUSO/OWL/AW-sized detector.
This possiblity was discussed in section \ref{sec:nuprimary}.

\subsection{Diffuse $\sim$GeV gamma-ray flux}
An upper limit on the diffuse gamma-ray flux 
between 30 MeV and 100 GeV 
has been published by the EGRET experiment of the now defunct
Gamma Ray Observatory \cite{EGRET}.  
This limit has serious
discriminatory power for models where SMPs or TDs or
extremely boosted massive particles decay to quark-antiquark jets
which then hadronization to produce the EECRs \cite{DiffusePho}.
This is because QCD jets via $\pi^0$ production and 
decay produce very high energy
gammas which initiate an electromagnetic cascade on the cosmic radio,
microwave, infrared, and magnetic field backgrounds.  
For models with jet-production distributed over
cosmic distances, such as some TD models and the $Z$-burst model
with a homogeneous distribution of relic neutrino targets,
the cascade has the distance to reach completion;
the end result is gamma power (energy/time) in the EGRET range
roughly an order of magnitude below
the total power of the original sources,
and comparable to the power in EECR neutrinos.  
Such models are disfavored.  
More local models, such as SMPs bound to our halo,
and the $Z$-burst model with a local over-density of relic neutrinos,
are not impacted at present, but may be tested with GLAST, 
the next generation gamma-ray observatory.

\section{CONCLUSIONS}
The ultimate explanation for the puzzles in EECRs
will provide a surprise at a minimum,
and possibly radically new physics at a maximum.
The extreme energies of events already observed cannot 
be approached by terrestrial accelerators.
Thus, there is ample motivation to build 
the next generation of CR detectors, Auger and Telescope Array,
and to plan even farther beyond for a teraton detector 
like EUSO, OWL and AirWatch.
At $10^{20}$~eV, AGASA provides about an event per year, and
HiRes about an event per month.  Auger and TA will see
two such events per week, 
while EUSO/OWL/AW has the potential to collect 
such an event every two hours.
As the data sample grows, statistical studies will
reveal signatures that discriminate among the many 
galactic and extragalactic sources 
so far proposed to resolve the EECR puzzles.

It is quite possible that neutrino primaries
are responsible for the EECRs.  If so, 
it appears that the weakly-interacting neutrino either grows a
very strong cross-section at $10^{20}$~eV, or 
it annhilates on the relic neutrinos left-over
from the hot phase of the Big Bang.
Another possibility is that free magnetic charges exist
and are the EECR primaries;
for magnetic monopoles, the Universe is transparent and 
cosmic magnetic fields provide a natural acceleration
mechanism.

Clearly, we live in exciting EECR times,
and we possess the technology to prove this is so.
The resolution of our puzzles is forthcoming, as 
on-going and future experiments will provide us with 
the statistics to discriminate among the many interesting models.
%
%\newpage
\section*{Acknowledgements}
We acknowledge fruitful collaboration with P. Biermann,
H. Goldberg, T. Kephart, H. Paes, and Dipthe Wick
on some of the material presented here;
and discussions with G. Farrar, G. Gelmini, A. Kusenko,
D. McKay, A. Olinto, J. Ralston, G. Sigl, F. Stecker, and 
E. Zas.
This work was supported in part by the
U.S. Department of Energy grant no.\ DE-FG05-85ER40226.
\end{document}